\def\bbbq{{\mathchoice 
{\setbox0=\hbox {$\displaystyle\rm Q$}\hbox
{\raise0.15\ht0\hbox to0pt{\kern0.4\wd0\vrule height0.8\ht0\hss}\box0}}
{\setbox0=\hbox {$\textstyle\rm Q$}\hbox
{\raise0.15\ht0\hbox to0pt{\kern0.4\wd0\vrule height0.8\ht0\hss}\box0}}
{\setbox0=\hbox {$\scriptstyle\rm Q$}\hbox
{\raise0.15\ht0\hbox to0pt{\kern0.4\wd0\vrule height0.7\ht0\hss}\box0}}
{\setbox0=\hbox {$\scriptscriptstyle\rm Q$}\hbox
{\raise0.15\ht0\hbox to0pt{\kern0.4\wd0\vrule height0.7\ht0\hss}\box0}}
}}
 
\def\bbbc{{\mathchoice 
{\setbox0=\hbox {$\displaystyle\rm C$}\hbox
{\hbox to0pt{\kern0.4\wd0\vrule height0.9\ht0\hss}\box0}}
{\setbox0=\hbox {$\textstyle\rm C$}\hbox
{\hbox to0pt{\kern0.4\wd0\vrule height0.9\ht0\hss}\box0}}
{\setbox0=\hbox {$\scriptstyle\rm C$}\hbox
{\hbox to0pt{\kern0.4\wd0\vrule height0.9\ht0\hss}\box0}}
{\setbox0=\hbox {$\scriptscriptstyle\rm C$}\hbox
{\hbox to0pt{\kern0.4\wd0\vrule height0.9\ht0\hss}\box0}}
}}
\font\fivesans=cmss10 at 5pt 
\font\sevensans=cmss10 at 7pt 
\font\tensans=cmss10   
\newfam\sansfam  
\textfont\sansfam=\tensans\scriptfont\sansfam=\sevensans  
\scriptscriptfont\sansfam=\fivesans  
\def\sans{\fam\sansfam\tensans} 
\def\bbbz{{\mathchoice {\hbox{$\sans\textstyle Z\kern-0.4em Z$}}  
{\hbox{$\sans\textstyle Z\kern-0.4em Z$}}  
{\hbox{$\sans\scriptstyle Z\kern-0.3em Z$}}  
{\hbox{$\sans\scriptscriptstyle Z\kern-0.2em Z$}}}} 

\def\slash#1{#1\kern-0.65em /}
\def\dirac{{\raise0.09em\hbox{/}}\kern-0.58em\partial}
\def\Dirac{{\raise0.09em\hbox{/}}\kern-0.69em D}
\def\kbar{{\mathchar'26\mkern-9muk}}  

\documentstyle[12pt]{article}
\begin{document}

\title{Noncommutative Kaluza-Klein Theory}

\author{J. Madore \\
        {\em LPTHE\thanks{Laboratoire associ\'e au CNRS, {\rm URA D0063}}, 
         Universit\'e de Paris-Sud}\\
        {\em B\^at. 211, F-91405 Orsay}
\and    
        J. Mourad \\
        {\em GPS, Universit\'e de Cergy Pontoise}\\
        {\em Site de St. Martin, F-95302 Cergy Pontoise}\\
       }

\date{January, 1996}

\maketitle

\abstract{Efforts have been made recently to reformulate traditional 
Kaluza-Klein theory by using a generalized definition of a
higher-dimensional extended space-time. Both electromagnetism and
gravity have been studied in this context. We review some of the models
which have been proposed, with a special effort to keep the mathematical
formalism to a very minimum.}


\vfill
\noindent
LPTHE Orsay 96/02
\medskip
\eject
\parskip 4pt plus2pt minus2pt

\section{Introduction and Motivation}\label{IntroductionandMotivation}

The simplest definition of noncommutative geometry is that it is a
geometry in which the coordinates do not commute. Perhaps not the
simplest but certainly the most familiar example is the quantized
version of a 2-dimensional phase space described by the `coordinates'
$p$ and $q$. This example has the advantage of illustrating what is for
us the essential interest of noncommutative geometry as expressed in the
Heisenberg uncertainty relations: the lack of a well-defined notion of a
point. `Noncommutative geometry is pointless geometry.'  The notion of a
point in space-time is often an unfortunate one. It is the possibility
in principle of being able to localize a particle at any length scale
which introduces the ultraviolet divergences of quantum field theory. It
would be interesting then to be able to modify the coordinates of space
and time so that they become noncommuting operators. By analogy with
quantum mechanics one could then expect points to be replaced by
elementary cells. This cellular structure would serve as an ultraviolet
cut-off similar to a lattice structure. The essential difference is that
it is possible in principle to introduce the cellular structure without
breaking Lorentz invariance.

When a physicist calculates a Feynman diagram he is forced to place a
cut-off $\Lambda$ on the momentum variables in the integrands. This
means that he renounces any interest in regions of space-time of
dimension less than $\Lambda^{-1}$. As $\Lambda$ becomes larger and
larger the forbidden region becomes smaller and smaller. The basic
assumption which we make is that this forbidden region cannot, not only
in practice but even in principle, become arbitrarily small. There is a
fundamental length scale below which the notion of a point makes no
sense. The simplest and most elegant, if not the only, way of doing this
in a Lorentz-invariant way is through the introduction of non-commuting
coordinates, exactly as in quantum mechanics.

To illustrate in more detail the analogy with quantum mechanics it is of
interest to examine the phase space of a classical particle moving in a
plane. In the language of quantum mechanics it is described by two
position operators $(q^1, q^2)$ and two momentum operators $(p_1, p_2)$.
These four operators all commute; they can be simultaneously
measured and the eigenvalues can be considered as the coordinates of the
points of a 4-dimensional space.  When the system is quantized they no
longer commute; they satisfy the canonical commutation relations
\begin{equation}
[q^1, p_1] = i\hbar, \qquad [q^2,p_2] = i\hbar.                \label{1.1}
\end{equation}
Because of this there is no longer a notion of a point in phase space
since one cannot measure simultaneously the position and momentum of a
particle to arbitrary precision.  However phase space can be thought of
as divided into cells of volume $(2\pi\hbar)^2$ and it is this cellular
structure which replaces the point structure. If the classical phase
space is of finite total volume there are a finite number of cells and
the quantum system has a finite number of possible states. A function on
phase space is defined then by a finite number of values.

In the presence of a magnetic field $H$ normal to the plane the momentum
operators must be further modified. They are replaced by the
minimally-coupled expressions and they also cease to commute:
\begin{equation}
[p_1, p_2] = i e H \hbar.                                      \label{1.2}
\end{equation}
This introduces a cellular structure in the momentum plane. It becomes
divided into `Landau cells' of area proportional to $eH \hbar$.
Consider in this case the divergent integral
\begin{equation}
I = \int {d^2 p \over p^2}.
\end{equation}
The commutation relations (\ref{1.2}) do not permit $p_1$ and $p_2$
simultaneously to take the eigenvalue zero and the operator $p^2$ is
bounded below by $eH\hbar$. The magnetic field acts as an infrared
cut-off. If one adds an {\it ad hoc} ultraviolet cut-off $\Lambda$ then
$p^2$ is bounded also from above and the integral becomes finite:
\begin{equation}
I \sim \log ({\Lambda^2 \over eH\hbar}).                   \label{1.3}
\end{equation}
If the position space were curved, with constant Gaussian curvature $K$,
one would have (\ref{1.2}) with the minimally-coupled expressions for
the momentum and with $eH\hbar$ replaced by $K\hbar^2$.  One would
obtain again an infrared regularization for $I$.

One can also suppose the coordinates of position space to be replaced by
two operators which do not commute:
\begin{equation}
[q^1, q^2] = i\kbar.                                              \label{1.4}  
\end{equation}
The constant $\kbar$ determines a new length scale which has no {\it a
priori} relation with $\hbar$ any more than (\ref{1.2}) has a relation
with (\ref{1.1}).  By the new uncertainty relation there is no longer a
notion of a point in position space since one cannot measure both
coordinates simultaneously but as before, position space can be thought
of as divided into cells. If we consider for example the divergent
integral $I$ and use the same logic that led to (\ref{1.3}) we find that
the commutation relations (\ref{1.4}) introduce an ultraviolet cut-off.
If we introduce also a constant Gaussian curvature and use the
equivalent of (\ref{1.2}) we have
\begin{equation}
I \sim \log (\kbar K).                                            \label{1.5}
\end{equation}
The integral has become completely regularized.

Although the ultimate ambition of noncommutative geometry (in physics)
is is to introduce a noncommutative version of space-time and to use it
to describe quantum gravity, we shall here address the much more modest
task of considering a modified version of Kaluza-Klein theory in which
the hidden `internal' space alone is described by a noncommutative
geometry.  The rational for this is the fact that the hidden dimensions,
if at all, are small. In the following Section we briefly recall the
basics of the standard version of Kaluza-Klein theory but in a notation
which makes it natural to pass to the modified version. This means above
all that we must introduce the notion of a differential in a rather
abstract way since later we shall be forced to take differentials of
matrices and other objects which in the usual sense of the word do not
possess derivatives.  This is the only technical part of the article
from a mathematical point of view. In
Section~\ref{NoncommutativeGeometry} we give a very rudimentary
introduction to some of the more elementary aspects of noncommutative
geometry, just sufficient so as to be able to pass in the following
Section~\ref{Kaluza-KleinTheoryRevisited} to the description of the
modified versions of Kaluza-Klein theory. This is the central section.
In it we describe models based on electromagnetism which purport to
describe various aspects of the standard model of electro-weak and
strong interactions. Only in the concluding
Section~\ref{NoncommutativeSpace-Time} shall we be interested in the
gravitational field.  We shall there describe what we consider might be
a relation between noncommutative geometry and classical and/or quantum
gravity. 

The subject has evolved considerably since a similar review was written
in 1988~\cite{Madore1988}. For an sampling of the early history of ideas
on the micro-texture of space-time we refer to Section~1.3 of the book
by Prugove\v cki~\cite{Prugovecki} as well as to the review articles by
Kragh \& Carazza~\cite{KraghCarazza} and Gibbs~\cite{Gibbs}.

\section{Kaluza-Klein Theory}\label{Kaluza-KleinTheory}

The question of whether or not space-time has really 4 dimensions, and
why, has been debated for many years. One of the first negative answers
was given by Kaluza \cite{Kaluza1921} and Klein \cite{Klein1926} in
their attempt to introduce extra dimensions in order to unify the
gravitational field with electromagnetism. Einstein \& Bergmann
\cite{EinsteinBergmann1938} suggested that at sufficiently small scales
what appears as a point will in fact be seen as a circle. Later, with
the advent of more elaborate gauge fields, it was proposed that this
internal space could be taken as a compact Lie group or something more
general.  The great disadvantage of these extra dimensions is that they
introduce even more divergences in the quantum theory and lead to an
infinite spectrum of new particles. In fact the structure is strongly
redundant and most of it has to be discarded. An associated problem is
that of localization.  We cannot, and indeed do not wish to have to,
address the question of the exact position of a particle in the extra
dimensions any more than we wish to localize it too exactly in ordinary
space-time. We shall take this as motivation for introducing in
Section~\ref{Kaluza-KleinTheoryRevisited} a modification of Kaluza-Klein
theory with an internal structure which is described by a noncommutative
geometry and in which the notion of a point does not exist. As
particular examples of such a geometry we shall choose only internal
structures which give rise to a finite spectrum of particles.

In its local aspects Kaluza-Klein theory is described by an extended
space-time of dimension $N = 4 + k$ with coordinates $x^i = (x^\alpha,
x^a)$. The $x^\alpha$ are the coordinates of space-time which, except
for the last section, one can consider to be Minkowski space; the $x^a$
are the coordinates of the internal space, which in this section will be
implicitly supposed to be space-like and `small'.  In
Section~\ref{Kaluza-KleinTheoryRevisited} it will be of purely algebraic
nature and not necessarily `small'.

One of the most important tools in differential geometry is the
differential of a function and the most important advance in
noncommutative geometry has been the realization by Connes
\cite{Connes1986, Connes1994} that the differential has a natural
extension to the noncommutative case. We shall define a differential by
a set of simple rules which makes it obvious that it is equivalent to a
derivative and ask the reader to believe that the rules have a rigorous
and natural mathematical foundation. He will see that they are quite
easy to manipulate in the simple noncommutative geometries we consider
in Section~\ref{NoncommutativeGeometry}

A 1-form is a covariant vector field $A_i$, which we shall write as 
$A = A_i dx^i$ using a set of basis elements $dx^i$. A 2-form is an
antisymmetric 2-index covariant tensor $F_{ij}$ which we shall write as
\begin{equation}
F = {1\over 2} F_{ij} dx^i dx^j
\end{equation}
using the product of the basis elements. This product is antisymmetric 
\begin{equation}
dx^i dx^j = - dx^j dx^i                                         \label{2.1}
\end{equation}
but otherwise has no relations.  Higher-order forms can be defined as
arbitrary linear combination of products of 1-forms. A $p$-form can be
thus written (locally) as
\begin{equation}
\alpha = {1\over p!} \alpha_{i_1 \cdots i_p} dx^{i_1} \cdots dx^{i_p}.
\end{equation}
The coefficients $\alpha_{i_1 \cdots i_p}$ are smooth functions and
completely antisymmetric in the $p$ indices. 

Let ${\cal A}$ be the set of complex-valued functions on the extended
space-time. Since the product of two functions can be defined, and is
independent of the order, ${\cal A}$ is a commutative algebra. We define
$\Omega^0({\cal A}) = {\cal A}$ and for each $p$ we write the vector
space of $p$-forms as $\Omega^p({\cal A})$. Each $\Omega^p({\cal A})$
depends obviously on the algebra ${\cal A}$ and, what is also obvious
and very important, it can be multiplied both from the left and the
right by the elements of ${\cal A}$.  It is easy to see that
$\Omega^p({\cal A}) = 0$ for all $p \geq n + 1$.  We define
$\Omega^*({\cal A})$ to be the set of all $\Omega^p({\cal A})$.  We have
seen that $\Omega^*({\cal A})$ has a product given by (\ref{2.1}).  It
is a graded commutative algebra.  It can be written as a sum
\begin{equation}
\Omega^*({\cal A}) = \Omega^+({\cal A}) \oplus \Omega^-({\cal A})\label{eo}
\end{equation}
of even forms and odd forms. The $A$ is an odd form and $F$ is even.
The algebra ${\cal A}$ is a subalgebra of $\Omega^+({\cal A})$.

Let $f$ be a function, an element of the algebra 
${\cal A} = \Omega^0({\cal A})$.  We define a map $d$ from
$\Omega^p({\cal A})$ into $\Omega^{p+1}({\cal A})$ by the rules
\begin{equation}
df = \partial_i f dx^i, \qquad d^2 = 0.                \label{differential}
\end{equation}
It takes odd (even) forms into even (odd) ones. From the rules we
find that
\begin{equation}
d A = d(A_i dx^i) 
= {1\over 2}(\partial_i A_j - \partial_j A_i) dx^i dx^j = F
\end{equation}
if we set
\begin{equation}
F_{ij} = \partial_i A_j - \partial_j A_i.
\end{equation}
From the second rule we have 
\begin{equation}
d F = 0.
\end{equation}
It is easy to see that if $\alpha$ is a $p$-form and $\beta$ is a 
$q$-form then
\begin{equation}
\alpha \beta = (-1)^{pq} \beta \alpha, \qquad
d(\alpha \beta) = (d \alpha) \beta + (-1)^p \alpha d\beta.
\end{equation}

The couple $(\Omega^*({\cal A}), d)$ is called a differential algebra or
a differential calculus over ${\cal A}$.  We shall see later that 
${\cal A}$ need not be commutative and $\Omega^*({\cal A})$ need not be
graded commutative. Over each algebra ${\cal A}$, be it commutative or
not, there can exist a multitude of differential calculi.  This fact
makes the noncommutative version of Kaluza-Klein theory richer than the
commutative version. As a simple example we define what is known as the
universal calculus $(\Omega_u^*({\cal A}), d_u)$ over the commutative
algebra of functions ${\cal A}$. We set, as always, 
$\Omega_u^0({\cal A}) = {\cal A}$ and for each $p \geq 1$ we define
$\Omega_u^p({\cal A})$ to be the set of $p$-point functions which vanish
when any two points coincide. It is obvious that 
$\Omega_u^p({\cal A}) \neq 0$ for all $p$, whatever $N$.  There is a map
$d$ from $\Omega_u^p({\cal A})$ into $\Omega_u^{p+1}({\cal A})$ given in
the lowest order by
\begin{equation}
(d_uf)(x^i, y^i) = f(y^i) - f(x^i).
\end{equation}
In higher orders it is given by a similar sort of alternating sum
defined so that $d_u^2 = 0$.  The algebra $\Omega_u^*({\cal A})$ is not
graded commutative.  It is defined for arbitrary functions, not
necessarily smooth, and it has a straightforward generalization for
arbitrary algebras, not necessarily commutative.

What distinguishes the usual differential calculus is the fact that it
is based on derivations. The derivative $\partial_i f$ of a smooth
function $f$ is a smooth function. We use the word derivation to
distinguish the map $\partial_i$ from the result of the map 
$\partial_i f$. A general derivation is a linear map $X$ from the
algebra into itself which satisfies the Leibniz rule: 
$X(fg) = (Xf)g + f(Xg)$. In the case we are presently considering a
derivation can always be written (locally) in terms of the basis 
$\partial_i$ as $X = X^i \partial_i$. Such is not always the case.
The relation between $d$ and $\partial_i$ is given by
\begin{equation}
df (\partial_i) = \partial_i f.                                \label{2.4}
\end{equation}
This equation has the same content as the first of (\ref{differential}).
One passes from one to the other by using the particular case
\begin{equation}
dx^i (\partial_j) = \delta^i_j.
\end{equation}
The basis $dx^i$ is said to be dual to the basis $\partial_i$.
The derivations form a vector space (the tangent space) at each point,
and (\ref{2.4}) defines $df$ as an element of the dual vector space
(the cotangent space) at the same point.  In the examples we consider in
Section~\ref{NoncommutativeGeometry} there are no points but the vector
spaces of derivations are still ordinary finite-dimensional vector
spaces.  Over an arbitrary algebra which has derivations one can always
define in exactly the same manner a differential calculus based on
derivations.  These algebras have thus at least two, quite different,
differential calculi, the universal one and the one based on the set of
all derivations.

To form tensors one must be able to define tensor products, for
example the tensor product 
$\Omega^1({\cal A}) \otimes_{\cal A} \Omega^1({\cal A})$ of 
$\Omega^1({\cal A})$ with itself.
We have here written in subscript the algebra ${\cal A}$. This
piece of notation indicates the fact that we identify $\xi f \otimes \eta$ 
with $\xi \otimes f \eta$ for every element $f$ of the algebra, a
technical detail which is important in the applications of
Section~\ref{Kaluza-KleinTheoryRevisited}. It means also that one must be
able to multiply the elements of $\Omega^1({\cal A})$ on the left
and on the right by the elements of the algebra ${\cal A}$. If ${\cal A}$ 
is commutative of course these two operations are equivalent. When 
${\cal A}$ is an algebra of functions this left/right linearity is
equivalent to the property of locality. It means that the product of a
function with a 1-form at a point is again a 1-form at the same point, a
property which distinguishes the ordinary product from other,
non-local, products such as the convolution. In the noncommutative case
there are no points and locality can not be defined; it is replaced by
the property of left and/or right linearity with respect to the algebra.

To define a metric and covariant derivatives on the extended space-time 
we set
\begin{equation}
\theta^i = dx^i                                                 \label{2.2a}
\end{equation}
in the absence of a gravitational field. We have then
\begin{equation}
d\theta^i = 0.                                                    \label{2.2}
\end{equation}
The extended Minkowski metric can be defined as the map
\begin{equation}
g(\theta^i \otimes \theta^j) = g^{ij}                             \label{2.3}
\end{equation}
which associates to each element $\theta^i \otimes \theta^j$ of the
tensor product $\Omega^1({\cal A}) \otimes_{\cal A} \Omega^1({\cal A})$
the contravariant components $g^{ij}$ of the (extended) Minkowski
metric.  There are of course several
other definitions of a metric which are equivalent in the case of
ordinary geometry but the one we have given has the advantage of an easy
extension to the noncommutative case. The map $g$ must be bilinear so
that we can define for arbitrary 1-forms $\xi = \xi_i \theta^i$ and
$\eta = \eta_i \theta^i$
\begin{equation}
g(\xi \otimes \eta) = \xi_i \eta_j g(\theta^i \otimes \theta^j) 
= \xi_i \eta_j g^{ij}.
\end{equation}

We introduce a gauge potential by first defining a covariant derivative.
Let $\psi$ be a complex-valued function which we shall consider as a
`spinor field' with no Dirac structure and let ${\cal H}$ be the space
of such `spinor fields'. A covariant derivative is a rule which
associates to each such $\psi$ in ${\cal H}$ a spinor-1-form $D\psi$. It
is a map
\begin{equation}
{\cal H} \buildrel D \over \rightarrow
\Omega^1({\cal A}) \otimes_{\cal A} {\cal H}                   \label{2.5}
\end{equation}
from ${\cal H}$ into the tensor product 
$\Omega^1({\cal A}) \otimes_{\cal A} {\cal H}$. In the absence of any
topological complications the function $\psi = 1$ is a spinor field and
we can define a covariant derivative by the rule
\begin{equation}
D 1 = A \otimes 1.
\end{equation}
The (local) gauge transformations are the complex-valued functions with
unit norm and so $A$ must be a 1-form with values in the Lie algebra of
the unitary group $U_1$, that is, the imaginary numbers.  An arbitrary
spinor field $\psi$ can always be written in the form 
$\psi = f \cdot 1 = 1 \cdot f$ where $f$ is an element of the algebra 
${\cal A}$. The extension to $\psi$ of the covariant derivative is given
by the Leibniz rule:
\begin{equation}
D \psi = df \otimes 1 + A \otimes f = d\psi \otimes 1 + A \otimes \psi,
\end{equation}
an equation which we shall simply write in the familiar form 
$D\psi = d\psi + A\psi$. Using the graded Leibniz rule
\begin{equation}
D (\alpha\psi) = d\alpha \otimes \psi + (-1)^p \alpha D \psi, \label{graded}
\end{equation}
the covariant derivative can be extended to higher-order forms and the
field strength $F$ defined by the equation
\begin{equation}
D^2 \psi = F \psi.
\end{equation}

To introduce the gravitational field it is always possible to maintain
(\ref{2.3}) but at the cost of abandoning (\ref{2.2}). This is known as
the moving-frame formalism. In the presence of gravity the $dx^i$ become
arbitrary 1-forms $\theta^i$. The differential $df$ can still be written
\begin{equation}
df = e_i f \theta^i                                          \label{Pfaff}
\end{equation}
in the form (\ref{differential}) provided one introduces modified 
derivations $e_i$.  We shall give explicitly expressions for such 
derivations in a noncommutative example in
Section~\ref{NoncommutativeGeometry}.  An equation 
\begin{equation}
df(e_i) = e_i f                                            \label{Pfaffder}
\end{equation}
equivalent to~(\ref{2.4}) can be written if one imposes the relations
\begin{equation}
\theta^i(e_j) = \delta^i_j.                  
\end{equation}
The $\theta^i$ are a (local) basis of the 1-forms dual to the
derivations $e_i$ exactly as the $dx^i$ are dual to the $\partial_i$.
Equation~(\ref{2.2}) must be replaced by the structure equations
\begin{equation}
d\theta^i = - {1\over 2} C^i{}_{jk} \theta^j \theta^k
\end{equation}
which express simply the fact that the differential of a 1-form is a
2-form and can be thus written out in terms of the (local) basis 
$\theta^i \theta^j$. The structure equations can normally not be written
globally and in the noncommutative case such equations do
not in general make sense because the differential forms need not have a
basis.

A covariant derivative is a rule which associates to each covariant 
vector $\xi$ a 2-index covariant tensor $D\xi$. It is a map 
\begin{equation}
\Omega^1({\cal A}) \buildrel D \over \rightarrow
\Omega^1({\cal A}) \otimes_{\cal A} \Omega^1({\cal A})         \label{2.6}
\end{equation}
from $\Omega^1({\cal A})$ into the tensor product
$\Omega^1({\cal A}) \otimes_{\cal A} \Omega^1({\cal A})$. 
On the extended space-time we can define a covariant derivative by the 
rule 
\begin{equation}
D\theta^i = -\Gamma^i{}_{jk} \theta^j \otimes \theta^k.
\end{equation}
The extension to arbitrary $\xi = \xi_i \theta^i$ is given by the
Leibniz rule:
\begin{equation}
D \xi = d\xi_i \otimes \theta^i 
- \xi_i \Gamma^i{}_{jk} \theta^j \otimes \theta^k.
\end{equation}
Using again a graded Leibniz rule, $D$ can be extended to higher-order
forms and the curvature $\Omega$ defined by the equation
\begin{equation}
D^2 \xi = - \Omega \xi = - \xi_i \Omega^i{}_j \otimes \theta^j.
\end{equation}
The curvature is the field strength of the gravitational field. The
minus sign is an historical convention. One can be write $\Omega^i{}_j$ 
in the form
\begin{equation}
\Omega^i{}_j = {1\over 2} R^i{}_{jkl} \theta^k \theta^l
\end{equation}
an equation which defines the components $R^i{}_{jkl}$ of the Riemann
tensor. 

The (local) gauge transformations are the functions with values in the
(local extended) Lorentz group. If one require that the torsion vanish and
that the covariant derivative be compatible with the metric then the
$\Gamma^i{}_{jk}$ are given uniquely in terms of the $C^i{}_{jk}$.

For a general introduction to Kaluza-Klein theory and to references to
the previous literature we refer to the review articles by Appelquist
{\it et al.}~\cite{Appelquist1987}, Bailin \& Love~\cite{BailinLove1987}
or Coquereaux \& Jadczyk~\cite{CoquereauxJadczyk1988}. Model building
using traditional Kaluza-Klein is developed, for example, by Kapetanakis
\& Zoupanos~\cite{Kapetanakis1992} and by Kubyshin {\it et
al.}~\cite{Kubyshin1989}.  On the extended space-time one can consider
gravity or one can consider, as a simpler problem, electromagnetism.
This was first done some time ago by Forg\'acs \&
Manton~\cite{ForgacsManton1980}, Manton~\cite{Manton1979}, Chapline \&
Manton~\cite{Chapline1980}, Fairlie~\cite{Fairlie1979} and Harnad {\it
et al.}~\cite{Harnad1980} The idea has a straightforward generalization
to noncommutative Kaluza-Klein theory which we shall discuss in
Section~\ref{ModelswithElectromagnetism}

\section{Noncommutative Geometry}\label{NoncommutativeGeometry}

The aim of noncommutative geometry is to reformulate as much as
possible the results of ordinary geometry in terms of an algebra of
functions and to generalize them to the case of a general noncommutative
(associative) algebra. The main notion which is lost when passing from
the commutative to the noncommutative case is that of a point.  The
original noncommutative geometry is the quantized phase space of
non-relativistic quantum mechanics.  In fact Dirac in his historical
papers in 1926 \cite{Dirac1926a, Dirac1926b} was aware of the
possibility of describing phase-space physics in terms of the quantum
analogue of the algebra of functions, which he called the quantum
algebra, and using the quantum analogue of the classical derivations,
which he called the quantum differentiations. And of course he was aware
of the absence of localization, expressed by the Heisenberg uncertainty
relation, as a central feature of these geometries. Inspired by work by
von Neumann \cite{vonNeumann1955} for several decades physicists studied
quantum mechanics and quantum field theory as well as classical and
quantum statistical physics giving prime importance to the algebra of
observables and considering the state vector as a secondary derived
object. This work has much in common with noncommutative geometry.  Only
recently has an equivalent of an exterior derivative been
introduced~\cite{Connes1988}.

The motivation for introducing noncommutative geometry in Kaluza-Klein
theory lies in the suggestion that space-time structure cannot be
adequately described by ordinary geometry to all length scales,
including those which are presumably relevant when considering hidden
dimensions. There is of course no reason to believe that the extra
structure can be adequately described by the simple matrix geometries
which we shall consider, although this seems well adapted to account for
the finite particle multiplets observed in nature. 

The simplest noncommutative algebras are the algebras $M_n$ of
$n \times n$ complex matrices.  Let $\lambda_a$ in $M_n$, for 
$1 \leq a \leq n^2-1$, be an antihermitian basis of the Lie algebra of
the special unitary group $SU_n$. The product $\lambda_a \lambda_b$ can
be written in the form
\begin{equation}
\lambda_a \lambda_b = {1\over 2} C^c{}_{ab} \lambda_c +
{1\over 2} D^c{}_{ab} \lambda_c - {1 \over n} g_{ab}.        \label{3.1.1} 
\end{equation}
The $g_{ab}$ are the components of the Killing metric; we shall use it
to raise and lower indices.  The $C^c{}_{ab}$ here are the structure
constants of the group $SU_n$ and $g_{cd}D^d{}_{ab}$ is trace-free and
symmetric in all pairs of indices.

We introduce derivations $e_a$ by
\begin{equation}
e_a f = [\lambda_a, f]
\end{equation}
for an arbitrary matrix $f$. It is an elementary fact of algebra that
any derivation $X$ of $M_n$ can be written as a linear combination 
$X = X^a e_a$ of the $e_a$ with the $X^a$ complex numbers. The complete
set of all derivations of $M_n$ will replace the space of all smooth
vector fields on the hidden part of extended space-time.

We define the algebra of forms $\Omega^*(M_n)$ over $M_n$ just as
we did in the commutative case. First we define $\Omega^0(M_n)$ to be
equal to $M_n$.  Then we use (\ref{Pfaffder}) to define $df$. 
This means in particular that
\begin{equation}
d\lambda^a(e_b) = [\lambda_b, \lambda^a ] = - C^a{}_{bc}\lambda^c.
                                                               \label{3.1.6}
\end{equation}
The algebra of forms $\Omega^*(M_n)$ and the extension of the
differential $d$ is defined exactly as in Section~\ref{Kaluza-KleinTheory}.
The big difference is that the algebra is not commutative and the
algebra of forms is not graded commutative. Graded commutativity can be
partially maintained however if instead of $d\lambda^a$ we use the 1-forms 
\begin{equation}
\theta^a = \lambda_b \lambda^a d\lambda^b.                      \label{3.1.9}
\end{equation}
These 1-forms have a special relation with the derivations. Instead of 
(\ref{3.1.6}) we have
\begin{equation}
\theta^a(e_b) = \delta^a_b                                      \label{3.1.7}
\end{equation}
a fact which makes calculations easier since
\begin{equation}
df = e_a f \theta^a                                           \label{3.1.8}
\end{equation}
as in (\ref{Pfaff}). From (\ref{3.1.7}) one can derive also the relations
\begin{equation}
\theta^a \theta^b = - \theta^b \theta^a, \qquad
f \theta^b = \theta^b f
\end{equation}
as well as
\begin{equation}
d\theta^a = 
-{1 \over 2} C^a{}_{bc} \, \theta^b \theta^c.                \label{3.1.10}
\end{equation}

From the generators $\theta^a$ we can construct a 1-form
\begin{equation}
\theta = - \lambda_a \theta^a.                                \label{3.1.11}
\end{equation}
Using $\theta$ we can rewrite (\ref{3.1.10}) as
\begin{equation}
d \theta + \theta^2 = 0.                                      \label{3.1.14}
\end{equation}
The interest in $\theta$ comes from the form
\begin{equation}
df = - [\theta,f]                                           \label{3.1.14a}
\end{equation}
for the differential of a matrix, an equation which follows directly
from (\ref{3.1.8})

One can use matrix algebras to construct examples of differential
calculi which have nothing to do with derivations.  Consider the algebra
$M_n$ graded as in supersymmetry with even and odd elements
and introduce a graded commutator between two matrices $\alpha$ and
$\beta$ as
\begin{equation}
[\alpha, \beta] = \alpha \beta - 
(-1)^{\vert \alpha \vert \vert \beta \vert} \beta \alpha
\end{equation}
where $\vert \alpha \vert$ is equal to 0 or 1 depending on whether
$\alpha$ is even or odd. One can define on $M_n$ a graded derivation
$\hat d$ by the formula
\begin{equation}
\hat d \alpha = - [ \eta , \alpha],                         \label{3.2.1}
\end{equation}
where $\eta$ is an arbitrary antihermitian odd element. Since $\eta$
anti-commutes with itself we find that $\hat d\eta = -2\eta^2$ and for
any $\alpha$ in $M_n$,
\begin{equation}
\hat d^2 \alpha = [ \eta^2, \alpha ].                         \label{3.2.2}
\end{equation}
The grading can be expressed as the direct sum 
$M_n = M_n^+ \oplus M_n^-$ of the even and odd elements of $M_n$.  This
decomposition is the analogue of (\ref{eo}).  If $n$ is even it is
possible to impose the condition
\begin{equation}
\eta^2 = - 1.                                                  \label{3.2.3}
\end{equation}
From (\ref{3.2.2}) we see that $\hat d^2 = 0$ and the map (\ref{3.2.1})
is a differential. In this case we shall write $\hat d = d$.  We see
that $\eta$ must satisfy
\begin{equation}
d\eta + \eta^2 = 1,                                           \label{3.2.3a}
\end{equation}
an equation which is to be compared with (\ref{3.1.14}).  If we 
define for all $p \geq 0$
\begin{equation}
\Omega^{2p}(M^+_n) = M^+_n, \qquad 
\Omega^{2p+1}(M^+_n) = M^-_n                                   \label{3.2.4}
\end{equation}
then we have defined a differential calculus over $M^+_n$. The 
differential algebra based on derivations can be imbedded in a larger
algebra such that a graded extension of (\ref{3.1.14a}) exists for 
all elements~\cite{Madore1995a}. In fact any differential calculus 
can be so extended.

As an example let $n=2$. To within a normalization the matrices
$\lambda_a$ can be chosen to be the Pauli matrices. We define
$\lambda_1$ and $\lambda_2$ to be odd and $\lambda_3$ and the identity
even. The most general possible form for $\eta$ is a linear combination
of $\lambda_1$ and $\lambda_2$ and it can be normalized so that
(\ref{3.2.3}) is satisfied.  Using $\Omega^*(M_2^+)$ one can construct a
differential calculus over the algebra of functions on a double-sheeted
space-time. This doubled-sheeted structure permits
one~\cite{ConnesLott1990} to introduce a description of parity breaking
in the weak interactions.

If $n$ is not even or, in general, if $\eta^2$ is not proportional to
the unit element of $M_n$ then $\hat d^2$ given by (\ref{3.2.2}) will
not vanish and $M_n$ will not be a differential algebra.  It is still
possible however to construct over $M_n^+$ a differential calculus
$\Omega^*(M_n^+)$ based on Formula~(\ref{3.2.1}). Essentially what one
does is just eliminate the elements which are the image of 
$\hat d^2$~\cite{ConnesLott1992}.

As an example let $n=3$. There is a grading defined by the
decomposition $3 = 2 + 1$ The most general possible form for $\eta$ is
\begin{equation}
\eta = \pmatrix{   0   &    0    & a_1 \cr 
                   0   &    0    & a_2  \cr 
                -a^*_1 & -a^*_2  &  0}.                      \label{3.2.10}
\end{equation}
For no values of the $a_i$ can (\ref{3.2.3}) be satisfied. The
general construction yields $\Omega^0(M_3^+) = M_3^+ = M_2 \times M_1$
and $\Omega^1(M_3^+) = M_3^-$ as in the previous example but after that
the elimination of elements which are the image of $\hat d^2$ reduces 
the dimensions. One finds $\Omega^2(M_3^+) = M_1$ and 
$\Omega^p(M_3^+) = 0$ for $p\geq 3$.

Consider the ordinary Dirac operator $\Dirac$ and let $\psi$ be a spinor
and $f$ a smooth function. It is straightforward to see that
\begin{equation}
e_a f i\gamma^a\psi = [\Dirac, f] \psi.                     \label{3.2.10a}
\end{equation}
If we make the replacement $\gamma^a \mapsto \theta^a$ the left-hand
side becomes equal to $i df \psi$. Because of the formal resemblance of
(\ref{3.1.14a}) and (\ref{3.2.1}) with this equation the matrices
$\theta$ and $\eta$ can be considered as generalized (antihermitian) 
Dirac operators. It is to be noticed that also $\Dirac^2 \neq 1$ 
and were one to use (\ref{3.2.10a}) to construct a differential
one would have also to eliminate unwanted terms. The problem here is
that $\theta^a \theta^b + \theta^b \theta^a = 0$ whereas
$\gamma^a \gamma^b + \gamma^b \gamma^a \neq 0$. If we consider the
algebra of functions ${\cal A}$ acting by multiplication on the Hilbert
space ${\cal H}$ of spinors then the ordinary differential calculus can
be described by the triple $({\cal A}, {\cal H}, \Dirac)$. As such it
can be generalized to the noncommutative case~\cite{Connes1986,
Connes1994}.  The triples for the examples above are 
$(M^+_2, \bbbc^2, \eta)$ and $(M^+_3, \bbbc^3, \eta)$ with $\eta$ in 
$M^-_2$ and $M^-_3$ respectively.

One can study `electromagnetism' on the algebras defined above, using
the differential calculi. Consider first the algebra $M_n$ with the
differential calculus based on derivations.  In the commutative case we
neglected the Dirac structure and considered a `spinor field' as an
element of the algebra of functions. Here we do the same. We identify a
`spinor field' $\psi$ as an element of the algebra $M_n$; it is a
`function' and it can be multiplied from the left by another arbitrary
`function' $f$.  A covariant derivative is a map of the form (\ref{2.5})
which for the same reasons we can write
\begin{equation}
D \psi = d\psi + \omega\psi.
\end{equation}
Using the graded Leibniz rule (\ref{graded}) the covariant derivative
can be extended to higher-order forms and the field strength $\Omega$
defined by the equation
\begin{equation}
D^2 \psi = \Omega \psi.
\end{equation}
By a simple calculation one finds that
\begin{equation}
\Omega = d\omega + \omega^2.
\end{equation}
The extra term arises because the algebra is noncommutative. Again by
strict analogy with the commutative case we define the gauge
transformations to be the group $U_n$ of unitary elements of $M_n$.  It
plays here the role of the {\it local} gauge transformations.  The
1-form $\omega$ must take its values in the Lie algebra of $U_n$ that
is, the set of antihermitian elements of $M_n$.  A particular choice of
$\omega$ is $\omega = \theta$. It is easy to verify that $\theta$ is
invariant under a gauge transformation. It makes sense then to decompose
$\omega$ as a sum $\omega = \theta + \phi$ and one sees that $\phi$
transforms under the adjoint action of the group $U_n$: 
$\phi \mapsto g^{-1} \phi g$. Expand $\phi = \phi_a \theta^a$. Then each
$\phi_a$ is a matrix. Using the identities (\ref{3.1.10}),
(\ref{3.1.14}) and (\ref{3.1.14a}) one sees that
\begin{equation}
\Omega = {1\over 2} \Omega_{ab} \theta^a \theta^b, \qquad
\Omega_{ab} = [\phi_a \phi_b] - C^c{}_{ab} \phi_c.
\end{equation}

One proceeds exactly in the same fashion with the algebra $M_n^+$ and
the differential calculus based on the generalized Dirac operator
$\eta$. One splits the gauge potential as a sum
\begin{equation}
\omega = \eta + \phi                                       \label{3.2.10b}
\end{equation}
and one finds, using the identity (\ref{3.2.1}) with $\hat d = d$ and
the identity (\ref{3.2.3a}), that
\begin{equation}
\Omega = 1 + \phi^2.                                        \label{3.2.11}
\end{equation}
Recall that the right-hand side is a 2-form.  In the two examples given
above, with $n=2$ and $n=3$, it can be identified as a real number.

One can also study `gravity' on $M_n$ using the differential calculus
based on derivations.  One defines a metric by the condition that the
$\theta^a$ be orthonormal with respect to the components of the
Killing metric:
\begin{equation}
g(\theta^a \otimes \theta^b) = g^{ab}.
\end{equation}
The unique metric-compatible torsion-free covariant derivative is given
by
\begin{equation}
D\theta^a = - {1\over 2} C^a{}_{bc}\theta^b \otimes \theta^c.   \label{3.4.3}
\end{equation}

On a matrix algebra there is a natural notion of integration defined by
the trace. For this and other further developments we refer to the
original literature. The basic texts on noncommutative geometry are the
books by Connes~\cite{Connes1986, Connes1994}. We refer also to a recent
physically oriented book~\cite{Madore1995a}.  The idea of using
derivations to define a differential calculus in the noncommutative case
was first considered by Dubois-Violette~\cite{Dubois-Violette1988}. The
1-forms $\theta^a$ were introduced and used to study noncommutative
gauge theory in a series of articles by Dubois-Violette {\it et al.}
\cite{Dubois-Violette1989a, Dubois-Violette1989b, Dubois-Violette1990a,
Dubois-Violette1990b, Dubois-Violette1991} The differential calculus
based on the generalized Dirac operator was introduced by
Connes~\cite{Connes1986, Connes1994}.  It was applied to matrices by
Connes \& Lott~\cite{ConnesLott1990, ConnesLott1992} and by Coquereaux
{\it et al.} \cite{Coquereaux1991}.  Other early references are the
articles by Connes~\cite{Connes1988}, Connes \&
Rieffel~\cite{ConnesRieffel1987} and by Coquereaux~\cite{Coquereaux1989}. 
It has been shown~\cite{MadoreMouradSitarz1995} that there is a sense in
which the calculus based on $M^+_2$ and the operator $\eta$ can be
considered as a singular deformation of the calculus using $M_2$ and its
derivations.  The introduction of `gravity' is much more difficult than
`electromagnetism' because of a technical problem coupled with the
structure of the 1-forms. If one compares (\ref{2.5}) with (\ref{2.6})
one sees that whereas one must be able to multiply elements of ${\cal
H}$ only from the left by elements of ${\cal A}$, one must be able to
multiply elements of $\Omega^1({\cal A})$ from the left and from the
right. In the noncommutative case these two actions are not the same.  A
solution to this problem has been suggested by Mourad~\cite{Mourad1995}
and developed by Dubois-Violette {\it et al.}~\cite{Dubois-Violette1995a,
Dubois-Violette1995b} and others~\cite{MadoreMassonMourad1995,
Georgelin1995}.

\section{Kaluza-Klein Theory Revisited}\label{Kaluza-KleinTheoryRevisited}

In traditional Kaluza-Klein theory the higher-order modes in the mode
expansion of the field variables in the coordinates of the internal
space are neglected, with the justification that they have all masses of
the order of the Planck mass and would not be of interest in
conventional physics. The alternative theory we here propose possesses
{\it ab initio} only a finite number of modes; there are no extraneous
modes to truncate.  We would like to suggest also that the
noncommutative version of Kaluza-Klein theory is more natural than the
traditional one in that a hand-waving argument can be given which allows
one to think of the extra algebraic structure as being due to quantum
fluctuations of the light-cone in ordinary 4-dimensional space-time. It
has been argued that this structure remains as a `classical shadow' of
the fluctuations, of which the noncommutative structure of space-time
itself is a higher-order correction.  Let $G$ be the gravitational
constant and set $\kbar = G \hbar$.  Let ${\cal A}_\kbar$ be the
regularized algebra of observables of quantum field theory, including
the regularizing gravitational field.  If one lets $\kbar \rightarrow 0$
then the algebra ${\cal A}_\kbar$ becomes completely singular by
assumption. It has no `classical' limit.  One can suppose however that
some subalgebra ${\cal Z}_\kbar \subset {\cal A}_\kbar$ remains regular
and has a commutative limit ${\cal Z}_0$ which one can identify as the
algebra of functions on space-time. We have supposed further that some
quasiclassical approximation exists which we can identify as a
Kaluza-Klein extension of space-time~\cite{MadoreMourad1995,
Kehagias1995}. The origin of this argument is the old idea, due to Pauli
and developed by Deser~\cite{Deser1957} and others~\cite{Isham1971},
that perturbative ultraviolet divergences will one day be regularized by
the gravitational field.

The version of Kaluza-Klein theory which we propose consists in
replacing the $k$ internal coordinates $x^a$ by generators of a
noncommutative algebra, for example the elements $\lambda_a$ introduced
in Section~\ref{NoncommutativeGeometry}. This means that the $k$ last
components $\theta^a$ of the $\theta^i$ defined in Equation~(\ref{2.2a})
must be replaced, for example by those defined by Equation~(\ref{3.1.9}).
A `moving frame' can be defined then by
\begin{equation}
\theta^i = (dx^\alpha, \lambda_b \lambda^a d\lambda^b).       \label{4.1}
\end{equation}
We have considered here the internal structure formally as being of
dimension $k = n^2 - 1$. This is however misleading since $n^2 - 1$ is
the dimension of all the `vector fields' on the algebraic structure, not
the dimension of the tangent space at one point.

If the geometry is one of those based on the generalized Dirac operator
$\eta$ then the more abstract notation must be used since there is no
basis $\theta^a$ and the total gauge potential $\omega$ must be written
in the index-free notation. Using (\ref{3.2.10b}) one has
\begin{equation}
\omega = A + \eta + \phi                                        \label{4.2}
\end{equation}
and one calculates the total curvature or field strength using the
identity (\ref{3.2.3a}).  Otherwise the development proceeds very much
as in traditional electromagnetism. Equation~(\ref{2.2}) for the
$\theta^a$ must be replaced by (\ref{3.1.10}) since the internal `space'
is `curved'.  The integral over the internal space becomes a trace over
the algebraic factor.  As we have already mentioned there are two
natural theories one can consider: the Maxwell-Dirac theory and the
Einstein-Dirac theory.

\subsection{Models with Electromagnetism}\label{ModelswithElectromagnetism}

Most of the efforts to introduce noncommutative geometry into particle
physics have been directed towards trying to find an appropriate
noncommutative generalization of the idea mentioned at the end of
Section~\ref{Kaluza-KleinTheory}. One studies electromagnetism on a
noncommutative extension of space-time and one calculates how the
particle and mass spectra vary as one varies the extra noncommutative
algebra and the associated differential calculus. Much ingenuity has
gone into these calculations which often involve very sophisticated
mathematics but which ultimately reduce to simple manipulations with
matrices. 

The idea then is, for example, to consider the electromagnetic gauge
potential $A = A_i \theta^i$ in an extended space-time but using the
basis (\ref{4.1}) instead of (\ref{2.2a}). Otherwise the formal
manipulations are the same. One arrives at a unification of Yang-Mills
and Higgs fields with the potential of the Higgs particle given by the
curvature of the covariant derivative in the algebraic `directions'. It
is quartic in the field variables since the Yang-Mills action is quartic
in the gauge potential. From the Expression~(\ref{3.2.11}) for the
curvature, for example, one can see the origin of the Higgs potential
normally introduced {\it ad hoc} to cause spontaneous symmetry breaking.
Of course the differential calculus has in this case been chosen
appropriately.

The simplest and most intuitive models are those based on derivations,
introduced by Dubois-Violette {\it et al.}~\cite{Dubois-Violette1989a}
and extended~\cite{Dubois-Violette1989b, Dubois-Violette1990a,
Dubois-Violette1990b, Madore1991, Balakrishna1991, Madore1993} soon
after.  The models based on the generalized Dirac operator are less
rigid and can be chosen to coincide with the Standard Model. The first
example, constructed by Connes \& Lott~\cite{ConnesLott1990}, was based
on the differential calculus defined by Equation~(\ref{3.2.4}) for
$n=2$. The extension to $n=3$ was given by Connes \&
Lott~\cite{ConnesLott1992}. A different extension to $n=4$ was developed
concurrently by Coquereaux {\it et al.}~\cite{Coquereaux1991,
CoquereauxHausslingPapadopoulosScheck, CoquereauxHausslingScheck},
Scheck~\cite{Scheck}, Papadopoulos {\it et
al.}~\cite{PapadopoulosPlassScheck}.  Further developments were due to
Iochum \& Sch\"ucker~\cite{Iochum1994}, Papadopoulos \&
Plass~\cite{PapadopoulosPlass} and Dimakis \&
M\"uller-Hoissen~\cite{Dimakis1994}.  Several review articles have been
written of these models. We refer, for example to the articles by
Kastler~\cite{Kastler1993}, by V\'arilly \& Gracia-Bond\'\i
a~\cite{Varilly1993} and by Kastler {\it et
al.}~\cite{IochumKastlerSchuecker1994}. A comparison of the two methods
has been made by Dubois-Violette {\it et al.}~\cite{Dubois-Violette1991}
and others~\cite{MadoreMouradSitarz1995}.

The weak interactions violate parity and this fact must be included in a
realistic model. No derivation-based model with explicit parity
violation has been developed; the models mentioned above rely implicitly
on spontaneous parity-breaking mechanisms like the `see-saw' mechanism.
As we have already mentioned the double-sheeted structure of the
Dirac-based models lends itself more readily to the introduction of
explicit parity violation. We refer to Alvarez {\it et
al.}~\cite{Alvarez1995} for a discussion of anomalies in this context.

Particle physics at the scale of grand unification has been examined
from the point of view of noncommutative geometry by Chamseddine {\it et
al.}~\cite{ChamsFeldFroh1992, ChamsFeldFroh1993a, ChamsFroh1994}, Batakis
{\it et al.}~\cite{BatakisKehagiasZoupanos, BatakisKehagias} and
others~\cite{Madore1993, Okumura}. Supersymmetry has also been
included~\cite{CoqEspScheck, HauPapScheck1991, HauPapScheck1993}. In
fact as was pointed out by Hussain \& Thompson~\cite{HussainThompson}
the noncommutative models based on the differential
calculi~(\ref{3.2.4}) are similar in structure to a `supersymmetric'
model proposed by Dondi \& Jarvis some time ago~\cite{DondiJarvis}.
Somewhat within the same context a completely different point of view of
the role of noncommutative geometry has been given by Iochum {\it et
al.}~\cite{IochumKastlerSchuecker1995}.

\subsection{Models with Gravity}\label{ModelswithGravity}

Very few of the results of the preceding subsection can be developed
within the context of the Einstein-Dirac theory and none of them have as
yet any significance for particle physics. We refer simply to the
original literature.  Gravity was first introduced in the context of
noncommutative geometry by Dubois-Violette {\it et
al.}~\cite{Dubois-Violette1989b} and developed in subsequent
articles~\cite{Madore1990, MadoreMourad1993, MadoreMourad1995,
Kehagias1995, Hajac}. The definition of curvature remains a
problem~\cite{Dubois-Violette1995b} as is the choice of action
functional~\cite{Connes1988, KalauWalze, Hajac, AckermannTolksdorf}. A
parallel development which treats gravity as an ordinary gauge field is
due to Chamseddine {\it et al.}~\cite{ChamsFeldFroh1993b,
ChamsFrohGrand} and others~\cite{Sitarz1994}. The details are given in
the lecture by Chamseddine.

\section{Noncommutative Space-Time}
\label{NoncommutativeSpace-Time}

We saw in the Introduction that a field theory in a noncommutative
version of space-time would have no ultraviolet divergences because
there would be no points. We saw also that the ultimate use of
noncommutative geometry as far as we are concerned is to describe
quantum and/or classical gravity. In
Section~\ref{Kaluza-KleinTheoryRevisited} we mentioned the old idea that
ultraviolet divergences will one day be regularized by the gravitational
field. The bridge between these ideas is the idea that noncommutative
structure of space-time is due to the quantum fluctuations of the
gravitational field~\cite{Doplicher1995, MadoreMourad1995,
Kehagias1995}. The first mention of noncommuting coordinates in
space-time in order to eliminate divergences was made by Snyder in
1947~\cite{Snyder1947a, Snyder1947b}.  We refer also to the early
article by Hellund \& Tanaka~\cite{Hellund} and to the more recent
lecture notes by Bacry~\cite{Bacry1988}.  Although the position of a
particle has no longer a well-defined meaning one can require that the
Lorentz group act on the algebra.  This was in fact the point which
Snyder was the first to make and which distinguishes a noncommutative
structure from the lattices which had been previously considered to
represent the micro-texture of space-time.  The space-time looks then
like a solid which has a homogeneous distribution of dislocations but no
disclinations. We can pursue this solid-state analogy and think of the
ordinary Minkowski coordinates as macroscopic order parameters obtained
by coarse-graining over scales less than the fundamental scale.  They
break down and must be replaced by elements of the algebra when one
considers phenomena on these scales.

A simple model in two dimensions with euclidean signature has been
introduced~\cite{Madore1992a} and developed~\cite{Madore1992b,
GrosseMadore1992, Madore1995a, Grosse1993, Grosse1995}. Although too
simple to give much intuition about the `correct' procedure it is an
interesting example of the correlation between noncommutativity and
curvature. A model in arbitrary dimension but with euclidean
signature~\cite{Madore1995b} is still in a preliminary state as is an
example based on an extension of the quantum
plane~\cite{DimakisMadore1995}.

Quite generally one can address the question of how far it is possible
to transcribe all of space-time physics into the language of
noncommutative geometry. We have seen that a differential calculus can
be constructed over an arbitrary associative algebra. This would permit
the formulation of gauge theories in any geometry. In a less general
setting a sort of Dirac operator has been proposed and a generalized
integral~\cite{Connes1988}. A serious problem is that of quantization.
The Standard Model is defined by a classical action which is assumed to
contain implicitly all of high-energy physics. Quantum corrections are
obtained by a standard quantization procedure. This quantization
procedure has not been generalized to noncommutative models even in the
simplest cases.  The examples which have been used to propose classical
actions which might be relevant in high-energy physics all involve
simple matrix factors. They are quantized by first expanding the
noncommutative fields in terms of ordinary space-time components and
then quantizing the components. Under quantization the constraints on
the model which come from the noncommutative geometry are 
lost~\cite{Dubois-Violette1989b, Alvarez1994, Hanlon}.

To conclude we mention two other closely related approaches to the
problem of the quantization of the gravitational field. It can be argued
that since one has `quantized' space one should also `quantize' the
Lorentz group. This idea leads to the notion of `quantum spaces' and
`quantum groups'. They are described in some detail in the lectures by
Castellani and by Wess.  The theory of strings is based on the
idea that the coordinates of (extended) space-time are fields on the
world surfaces of string-like objects. When quantized they naturally
become noncommuting objects. Under certain circumstances they even 
`noncommute' classically~\cite{Abouelsaood}.

\section*{Acknowledgments} This review was written while the authors
were attending the 5th Hellenic School and Workshops on Elementary
Particle Physics. They would like to take this occasion to thank the
organizers for their hospitality. They would like also to thank A.
Kehagias, L. McCulloch, O. P\`ene and T. Sch\"ucker for helpful
comments.


\begin{thebibliography}{99}
\parskip 5pt plus2pt minus2pt


\bibitem{AckermannTolksdorf}
Ackermann Thomas, Tolksdorf J\"urgen, ``A generalized Lichnerowicz
formula, the Wodzicki Residue and Gravity'', Preprint CPT-94/P.3106.

\bibitem{Alvarez1994}
Alvarez E., Gracia-Bond\'\i a J.M., Mart\'\i n C.P., ``A renormalization
group analysis of the NCG constraints $m_{\rm top} = 2 m_W$,
$m_{\rm Higgs} = 3.14 m_W$'', Phys. Lett. {\bf B329} (1994) 259.

\bibitem{Alvarez1995}
---, ``Anomaly cancellation and the gauge group of the Standard 
Model in NCG'', 
Phys. Lett. {\bf B364} (1995) 33.

\bibitem{Appelquist1987}
Appelquist T., Chodos A., Freund P.G.O., ``Modern
Kaluza-Klein Theory and Applications'', 
Benjamin-Cummings, 1987.

\bibitem{Abouelsaood}
Abouelsaood A., Callan C.G., Nappi C.R., Yost S.A., ``Open Strings in
Background Gauge Fields'', Nucl. Phys. {\bf B280} (1987) 599.

\bibitem{Bacry1988}
Bacry H., ``Localizability and Space in Quantum Physics'',
Lecture Notes in Physics No. 308, Springer Verlag, 1988.

\bibitem{BailinLove1987}
Bailin D., Love A., ``Kaluza-Klein theories'', 
Rep. Prog. Phys. {\bf 50} (1987) 1087.

\bibitem{Balakrishna1991}
Balakrishna B.S., G\"ursey F., Wali K.C., ``Towards a Unified
Treatment of Yang-Mills and Higgs Fields'', 
Phys. Rev. {\bf D44} (1991) 3313.

\bibitem{BatakisKehagias}
Batakis N.A., Kehagias A.A., ``On the construction of 
$SU(n) \times U(1)$ models in a non-commutative geometry setting'',
Class. Quantum Grav. {\bf 11} (1994) 2627.

\bibitem{BatakisKehagiasZoupanos}
Batakis N.A., Kehagias A.A., Zoupanos G., ``Structure and spontaneous 
symmetry breaking of a gauge theory based on $SU(5\vert 1)$'',
Phys. Lett. {\bf B315} (1993) 319

\bibitem{ChamsFroh1994}
Chamseddine A.H., Fr\"ohlich J., ``$SO(10)$ unification in
noncommutative geometry'', Phys. Rev. {\bf D50} (1994) 2893.

\bibitem{ChamsFeldFroh1992}
Chamseddine A.H., Felder G., Fr\"ohlich J., ``Unified gauge theories in
non-commutative geometry'', Phys. Lett. {\bf B296} (1992) 109.

\bibitem{ChamsFeldFroh1993a}
---, ``Grand unification in non-commutative geometry'', 
Nucl. Phys. {\bf B395} (1993) 672.

\bibitem{ChamsFeldFroh1993b}
---, ``Gravity in Non-Commutative Geometry'', 
Commun. Math. Phys. {\bf 155} (1993) 205.

\bibitem{ChamsFrohGrand}
Chamseddine A.H., Fr\"ohlich J., Grandjean O., ``The gravitational sector
in the Connes-Lott formulation of the standard model'',
J. Math. Phys. {\bf 36} (1995) 6255.

\bibitem{Chapline1980}
Chapline G., Manton N. S., ``The Geometrical Significance of Certain
Higgs Potentials: An approach to grand unification'', 
Nucl. Phys. {\bf B184} (1980) 391.

\bibitem{Connes1986}
Connes A., ``Non-Commutative Differential Geometry'', 
Publications of the Inst. des Hautes Etudes Scientifiques. {\bf 62} 
(1986) 257.

\bibitem{Connes1988}
---, ``The Action Functional in Non-Commutative Geometry'',
Commun. Math. Phys. {\bf 117} (1988) 673.

\bibitem{Connes1994}
---, ``Noncommutative Geometry'', 
Academic Press, 1994.

\bibitem{ConnesLott1990}
Connes A., Lott J., ``Particle Models and Noncommutative Geometry'',
in `Recent Advances in Field Theory', 
Nucl. Phys. Proc. Suppl. {\bf B18} (1990) 29.

\bibitem{ConnesLott1992}
---, ``The metric aspect of non-commutative geometry'', 
Proceedings of the 1991 Carg\`ese Summer School, Plenum Press, 1992.

\bibitem{ConnesRieffel1987}
Connes A., Rieffel M.A., ``Yang-Mills for non-commutative two-tori'',
Contemp. Math. {\bf 62} (1987) 237.

\bibitem{Coquereaux1989}
Coquereaux R., ``Noncommutative geometry and theoretical physics'', 
J. Geom. Phys. {\bf 6} (1989) 425.

\bibitem{CoquereauxJadczyk1988}
Coquereaux R., Jadczyk A., ``Riemannian Geometry Fiber Bundles 
Kaluza-Klein Theories and all that....'', 
World Scientific Lecture Notes in
Physics {\bf 16} (1988), World Scientific, Singapore.

\bibitem{Coquereaux1991}
Coquereaux R., Esposito-Far\`ese G., Vaillant G. 1991, ``Higgs Fields as
Yang-Mills Fields and Discrete Symmetries'', 
Nucl. Phys. {\bf B353} (1991) 689.

\bibitem{CoqEspScheck}
Coquereaux R., Esposito-Far\`ese G., Scheck F.,
``Non-Commutative Geometry and Graded Algebras in Electroweak Interactions'',
J. Mod. Phys. {\bf A7} (1992) 6555.

\bibitem{CoquereauxHausslingScheck}
Coquereaux R., H\"au\ss ling R., Scheck F., ``Algebraic connections on 
parallel universes'', 
Int. J. Mod. Phys. {\bf A10} (1995) 89. 

\bibitem{CoquereauxHausslingPapadopoulosScheck}
Coquereaux R., H\"au\ss ling R., Papadopoulos N.A., Scheck F.,
``Generalized gauge transformations and hidden symmetry in the 
standard model'',
Int. J. Mod. Phys. {\bf A7} (1992) 2809.

\bibitem{Deser1957}
Deser S., ``General Relativity and the Divergence Problem in Quantum
Field Theory'',
Rev. Mod. Phys. {\bf 29} (1957) 417.

\bibitem{Dimakis1994}
Dimakis A., M\"uller-Hoissen F., ``Discrete Differential Calculus, 
Graphs, Topologies and Gauge Theory'', 
J. Math. Phys. {\bf 35} (1994) 6703.

\bibitem{DimakisMadore1995}
Dimakis A., Madore J., ``Differential Calculi and Linear Connections'', 
Preprint, LPTHE Orsay, 95/79, q-alg/9601023.

\bibitem{Dirac1926a}
Dirac P.A.M., ``The Fundamental Equations of Quantum Mechanics'', 
Proc. Roy. Soc. {\bf A109} (1926) 642.

\bibitem{Dirac1926b}
---, ``On Quantum Algebras'', 
Proc. Camb. Phil. Soc. {\bf 23} (1926) 412.

\bibitem{DondiJarvis}
Dondi P.H., Jarvis P.D., ``A supersymmetric Weinberg-Salam Model'',
Phys. Lett. {\bf B84} (1979) 75.

\bibitem{Doplicher1995}
Doplicher S., Fredenhagen K., Roberts J.E., ``The Quantum Structure
of Spacetime at the Planck Scale and Quantum Fields'', 
Commun. Math. Phys. {\bf 172} (1995) 187.

\bibitem{Dubois-Violette1988}
Dubois-Violette M. 1988, ``D\'erivations et calcul diff\'erentiel 
non-commutatif'',
C. R. Acad. Sci. Paris {\bf 307} S\'erie I (1988) 403.

\bibitem{Dubois-Violette1989a}
Dubois-Violette M., Kerner R., Madore J., ``Gauge bosons in a
noncommutative geometry'', 
Phys. Lett. {\bf B217} (1989) 485. 

\bibitem{Dubois-Violette1989b}
---, ``Classical bosons in a noncommutative geometry'', 
Class. Quant. Grav. {\bf 6} (1989) 1709.

\bibitem{Dubois-Violette1990a}
---, ``Noncommutative differential geometry of matrix algebras'',
J. Math. Phys. {\bf 31} (1990) 316.

\bibitem{Dubois-Violette1990b}
---, ``Noncommutative differential geometry and new models of gauge theory'', 
J. Math. Phys. {\bf 31} (1990) 323.

\bibitem{Dubois-Violette1991}
---, ``Super Matrix Geometry'', 
Class. Quant. Grav. {\bf 8} (1991) 1077.

\bibitem{Dubois-Violette1995a}
Dubois-Violette M., Madore J., Masson T., Mourad J. 1995, ``Linear
Connections on the Quantum Plane'', 
Lett. Math. Phys. {\bf 35} (1995) 351.

\bibitem{Dubois-Violette1995b}
---, ``On Riemann Curvature in Noncommutative Geometry'',
Preprint LPTHE Orsay 95/63, q-alg/9512004.

\bibitem{EinsteinBergmann1938}
Einstein A., Bergmann P., ``On a Generalization of Kaluza's Theory
of Electricity'', 
Ann. of Math. {\bf 39} (1938) 683.

\bibitem{Fairlie1979}
Fairlie, D.B., ``The Interpretation of Higgs Fields as Yang Mills Fields'', 
Geometrical and Topological Methods in Gauge Theories, Proceedings,
Montreal, Lecture Notes in Physics {\bf 129} (1979) 45.

\bibitem{ForgacsManton1980}
Forg\'acs P., Manton N.S., ``Space-Time Symmetries in Gauge Theories'',
Commun. Math. Phys. {\bf 72} (1980) 15.

\bibitem{Gibbs}
Gibbs Phil, ``The Small Scale Structure of Space-Time: A Bibliographical
Review'', Preprint, hep-th/9506171.

\bibitem{Georgelin1995}
Georgelin Y., Madore J., Masson T., Mourad J., 
``On the non-commutative Riemannian geometry of $GL_{q}(n)$'',
Preprint LPTHE Orsay 95/51.

\bibitem{GrosseMadore1992}
Grosse H., Madore J., ``A Noncommutative Version of the Schwinger Model'',
Phys. Lett. {\bf B283} (1992) 218.

\bibitem{Grosse1993}
Grosse H., Pre\v snajder P., ``The Construction of Noncommutative
Manifolds Using Coherent States'',
Lett. in Math. Phys. {\bf 28} (1993) 239.

\bibitem{Grosse1995}
Grosse H., Klim\v c\'\i k C., Pre\v snajder P., ``Field Theory on a
Supersymmetric Lattice'', Preprint, CERN-TH/95-195.

\bibitem{Hajac}
Hajac Pietr M., ``The Einstein Action for Algebras of Matrix Valued
Functions - Toy Models'', ICTP Preprint, IC/95/352.

\bibitem{Hanlon}
Hanlon B.E., Joshi G.C., ``BRS symmetry in Connes' non-commutative 
geometry'', J. Phys. A {\bf 28} (1995) 2889.

\bibitem{Harnad1980}
Harnad J., Shnider S., Tafel J., ``Group actions on principal bundles
and dimensional reduction'', 
Lett. in Math. Phys. {\bf 4} (1980) 107.

\bibitem{HauPapScheck1991}
H\"au\ss ling R., Papadopoulos N.A., Scheck F., ``$SU(2|1)$ Symmetry,
Algebraic Superconnections and Generalized Theory of Electroweak 
Interactions'', 
Phys. Lett. {\bf B260} (1991) 125.

\bibitem{HauPapScheck1993}
-- , ``Supersymmetry in the Standard Model of Electroweak Interactions'',
Phys. Lett. {\bf B303} (1993) 265.

\bibitem{Hellund}
Hellund Emile J., Tanaka Katsumi, ``Quantized Space-Time''
Phys. Rev. {\bf 94} (1954) 192.

\bibitem{HussainThompson}
Hussain F., Thompson G. 1991, ``Non-commutative geometry and
supersymmetry'', 
Phys. Lett. {\bf B260} (1991) 359.

\bibitem{Iochum1994}
Iochum B., Sch\"ucker T., ``A Left-Right Symmetric Model \`a la 
Connes-Lott'',  Lett. Math. Phys. {\bf 32} (1994) 153.

\bibitem{IochumKastlerSchuecker1994}
Iochum B., Kastler D., Sch\"ucker T., ``Fuzzy Mass Relations for the Higgs'',
J. Math. Phys. (1996) to appear.

\bibitem{IochumKastlerSchuecker1995}
---, ``Riemannian and Non-commutative Geometry in Physics'', 
Preprint CPT-95/P.3260, hep-th/9511011.

\bibitem{Isham1971}
Isham C.J., Salam A., Strathdee J., ``Infinity Suppression in 
Gravity-Modified Quantum Electrodynamics'', 
Phys. Rev. {\bf D3} (1971) 1805.

\bibitem{KalauWalze}
Kalau W., Walze M. ``Gravity, Non-Commutative Geometry and the Wodzicki
Residue'', Jour. Geom. and Phys. {\bf 16} (1995) 327.

\bibitem{Kaluza1921}
Kaluza Th., ``Zum Unit\"atsproblem der Physik'', 
Sitz. Preuss. Akad. Wiss. {\bf K1} (1921) 966.

\bibitem{Kapetanakis1992}
Kapetanakis D., Zoupanos G., ``Coset-Space Dimensional Reduction of
Gauge Theories'',
Phys. Rep. {\bf 219} (1992) 1.

\bibitem{Kastler1993}
Kastler D., ``A detailed account of Alain Connes' version of the
standard model in non-commutative geometry'',
Rev. Math. Phys. {\bf 5} (1993) 477.

\bibitem{Kehagias1995}
Kehagias A., Madore J., Mourad J., Zoupanos G., ``Linear Connections in
Extended Space-Time'',
J. Math. Phys. {\bf 36} (1995) 5855.

\bibitem{Klein1926}
Klein O., ``Quantentheorie und funfdimensionaler Relativit\"atstheorie'', 
Z. Phys. {\bf 37} (1926) 895.

\bibitem{KraghCarazza}
Kragh H. Carazza B. ``From Time Atoms to Space-Time Quantization: the
Idea of Discrete Time, ca 1925-1936'', Stud. Hist. Phil. Sci. {\bf 25}
(1994) 437. 

\bibitem{Kubyshin1989}
Kubyshin Y.A., Mour\~ao J.M., Rudolph G., Volobujev I.P., ``Dimensional 
Reduction of Gauge Theories, Spontaneous Compactification and Model 
Building'', Lecture Notes in Physics, {\bf349} (1989), Springer Verlag.

\bibitem{Madore1988}
Madore J., ``Kaluza-Klein Aspects of Noncommutative Geometry'',
Proceedings of the XVII International Conference on Differential
Geometric Methods in Theoretical Physics, Chester, August, 1988.

\bibitem{Madore1990}
---, ``Modification of Kaluza-Klein Theory'',
Phys. Rev. {\bf D41} (1990) 3709.

\bibitem{Madore1991}
---, ``Algebraic Structure and Particle Spectra'', 
Int. J. Mod. Phys. {\bf A6} (1991)1287.

\bibitem{Madore1992a}
---, ``The Fuzzy Sphere'', 
Class. Quant. Grav. {\bf 9} (1992) 69.

\bibitem{Madore1992b}
---, ``Fuzzy Physics'', 
Annals of Physics {\bf 219} (1992) 187.

\bibitem{Madore1993}
---, ``On a Lepton-Quark Duality'', 
Phys. Lett. {\bf B305} (1993) 84.

\bibitem{Madore1995a}
---, ``An Introduction to Noncommutative Differential
Geometry and its Physical Applications'', 
Camb. Univ. Press, 1995. 

\bibitem{Madore1995b}
---, ``Linear Connections on Fuzzy Manifolds'',
Preprint LPTHE Orsay 95/42, ESI Vienna 235, hep-th/9506183.

\bibitem{MadoreMourad1993}
Madore J., Mourad J., ``Algebraic-Kaluza-Klein Cosmology'', 
Class. Quant. Grav. {\bf 10} (1993) 2157.

\bibitem{MadoreMourad1995}
---, ``On the Origin of Kaluza-Klein Structure'', 
Phys. Lett. {\bf B359} (1995) 43.

\bibitem{MadoreMassonMourad1995}
Madore J., Masson T., Mourad J., ``Linear Connections on Matrix 
Geometries'', Class. Quant. Grav. {\bf 12} (1995) 1429.

\bibitem{MadoreMouradSitarz1995}
Madore J., Mourad J., Sitarz A. ``Deformations of Differential Calculi'', 
Preprint, LPTHE Orsay 95/75, hep-th/9601120.

\bibitem{Manton1979}
Manton N.S., ``A New Six-Dimensional Approach to the 
Wein\-berg-Salam Model'',  
Nucl. Phys. {\bf B158} (1979) 141.

\bibitem{Mourad1995}
Mourad. J., ``Linear Connections in Non-Commutative Geometry'',
Class. Quant. Grav. {\bf 12} (1995) 965.

\bibitem{Okumura}
Okumura Y., ``$SO(10)$ Grand Unified Theory in Non-Commutative
Differential Geometry on the Discrete Space $M_4 \times \bbbz_N$'',
Prog. Theor. Phys. {\bf 94} (1995) 607.

\bibitem{PapadopoulosPlass}
Papadopoulos N.A., Plass J., ``Natural extensions of the
Connes-Lott model and comparison with the Marseille-Mainz model'',
Mainz Preprint MZ-TH/95-11.

\bibitem{PapadopoulosPlassScheck}
Papadopoulos N.A., Plass J., Scheck F., ``Models of electroweak
interactions in non-commutative geometry: a comparison'', 
Phys. Lett. {\bf B324} (1994) 380.

\bibitem{Prugovecki}
Prugove\v cki E., ``Principles of Quantum General Relativity'', 
World Scientific, Singapore, 1995.

\bibitem{Scheck}
Scheck F., ``Anomalies, Weinberg angle and a non-commutative
description of the standard model'', 
Phys. Lett. {\bf B284} (1992) 303.

\bibitem{Sitarz1994}
Sitarz A., ``Gravity from non-commutative geometry'', 
Class. Quant. Grav. {\bf 11} (1994) 2127.  

\bibitem{Snyder1947a}
Snyder H.S., ``Quantized Space-Time'', 
Phys. Rev. {\bf 71} (1947) 38.

\bibitem{Snyder1947b}
---, ``The Electromagnetic Field in Quantized Space-Time'', 
Phys. Rev. {\bf 72} (1947) 68.

\bibitem{Varilly1993}
V\'arilly, J.C., Gracia-Bond\'\i a J.M., ``Connes' noncommutative
differential geometry and the Standard Model'', 
Jour. of Geom. and Phys. {\bf 12} (1993) 223.

\bibitem{vonNeumann1955}
von Neumann J., ``Mathematical Foundations of Quantum Mechanics'',
Princeton Univ. Press, 1955.


\end{thebibliography}
\end{document}